\documentclass[12pt]{article}
\usepackage{amssymb}
\usepackage[dvips]{epsfig}
\usepackage[utf8]{inputenc}
\usepackage{graphicx}
\usepackage{amsmath} 
\usepackage{bbm}
\usepackage{bm}

\usepackage{array}

\usepackage{hyperref}
\usepackage{longtable}

\def \del{\partial}

\newcommand{\dlb}{\delta_{\bar{\Lambda}}}
\newcommand{\hmn}{h_{\mu\nu\rho}}
\newcommand{\vf}{\varphi}

\newcommand{\be}{\begin{equation}}
\newcommand{\ee}{\end{equation}}
\def\bea{\begin{eqnarray}}
\def\eea{\end{eqnarray}}
\newcommand{\bn}{\begin{eqnarray}}
\newcommand{\en}{\end{eqnarray}}
\newcommand{\nn}{\nonumber}
\newcommand{\no}{\noindent}

\newcommand{\tchi}{\tilde{\chi}}

\newcommand{\p}{\partial}
\def\bea{\begin{eqnarray}}
\def\eea{\end{eqnarray}}
\newcommand{\beq}{\begin{eqnarray}}
\newcommand{\eeq}{\end{eqnarray}}

\begin{document}

\title{\textbf{Equivalence of spin-2 and spin-3 models invariant under transverse diffeomorphisms and the tensionless limit of string theory}}

	\author{ R. Schimidt Bittencourt \footnote{raphael.schimidt@unesp.br}, D. Dalmazi \footnote{denis.dalmazi@unesp.br}, B. dos S. Martins \footnote{bruno.s.martins@unesp.br}, E.L. Mendonça \footnote{elias.leite@unesp.br}\\
		\textit{{UNESP - Campus de Guaratinguetá - DFI }}} 
	\date{\today}
	\maketitle

\begin{abstract}

Here we investigate a general class of massless local theories of spin-2 and spin-3, both invariant under generalized transverse diffeomorphisms (TDiff).
We identify the ghost-free region in their parameter space and show the relationship of those models with the ``doublet'' action stemming from the tensionless limit of the open bosonic string field theory (for symmetric tensors). The connection is implemented via a nonlocal field redefinition which introduces a Stueckelberg-like field of rank-0 (rank-1) for the spin-2 (spin-3) case. An apparent
mismatch between most TDiff models and the ``doublet'' action has led us
to prove a nontrivial equivalence among TDiff models, thereby restoring
consistency. Any point in the parameter subspace of ghost-free TDiff models is equivalent to any other one within that subspace. In particular, they are all physically equivalent to their simplest versions known as Maxwell-like models.  So, the physical TDiff models seem to differ from each other by a BRST cohomologically trivial term. 
 

\end{abstract}

\newpage


\section{Introduction}

In recent years, there has been growing interest in the
connection between higher spin (HS) particles and string theory, as seen
in the review works  \cite{snow,rt,sagnotti_notes,st_r} and the book \cite{book}. An interesting point concerns the formulation of interacting theories for massless HS particles. On one hand, there are powerful No-Go theorems regarding the interaction of massless HS particles, see for instance \cite{porrati,bbs}, on the other hand massive HS particles in string theory can consistently interact and they
 become massless in the tensionless limit of the strings. 
 Moreover, the fact that HS models in their formulation in terms of symmetric tensors  makes use of constrained 
fields and constrained gauge symmetries \cite{fronsdal,sv}, while no such constraints seem to be natural in string theory is rather intriguing. Those issues will be addressed here specifically for spin-2 and spin-3 particles as reducible theories (TDiff models) in order to make contact with the spectrum of bosonic open-string field theory in the tensionless limit, which is also reducible.

First, we recall that
transverse diffeomorphisms (TDiff), namely, $\delta \, h_{\mu\nu} = \p_{\mu}\Lambda_{\nu}^T + \p_{\nu}\Lambda_{\mu}^T$ where  $\p^{\mu}\Lambda^T_{\mu}=0$ is the minimal symmetry  for the description of massless spin-2 particles, see \cite{van}. Such symmetry is enough to remove spin-1 ghosts as explained in detail in \cite{blas}. The TDiff model is reducible in general\footnote{As explained in \cite{blas} the TDiff spin-2 symmetry can be enlarged to Diff or to WTDiff (Weyl + TDiff), in both cases the model becomes irreducible. Higher spin ($s\ge 3$) models described by arbitrary rank-s symmetric fields can also be invariant under WTDiff and become irreducible \cite{sv}.}, it describes spin-2 and spin-0 massless particles, the latter ones may be physical or not, depending upon the free parameters of the model.

The simplest sub case of the TDiff models is known as Maxwell-like model \cite{cf}. It has been defined for arbitrary integer spin in \cite{cf}, in flat and AdS backgrounds, not only for symmetric but also for mixed symmetry tensors. For rank-s symmetric fields in four dimensional flat backgrounds it describes massless particles of spin $s,s-2, s-4, \cdots 0 $ or $1$ according to even or odd $s$ respectively.  This is exactly the same spectrum of the tensionless limit of the  open bosonic string field theory in the first Regge trajectory \cite{ouvry,beng,ht}. Due to the redefinition of the Virasoro generators and the tensionless limit ($\alpha'\to \infty$), the redefined Virasoro algebra has no central charge and no critical dimension is required. 

It is known that the tensionless limit of the mentioned string field theory leads to the so called triplet action, see \cite{st} and also \cite{ouvry,beng,ht}, which contains, in the first Regge trajectory, three symmetric fields $(\varphi,C,D)$ of rank $s,s-1$ and $s-2$ respectively. In order to connect the triplet action with the generalized TDiff models one first eliminates the rank $(s-1)$ C-field algebraically and derive a doublet action \cite{st} in terms of  $(\varphi,D)$. The connection with the TDiff Maxwell-like models can proceed further by eliminating the D-field via gauge fixing at action level or, on the singlet TDiff side,  by the introduction of a rank-$(s-1)$ Stueckelberg field which gives rise to the $D$-field afterwards,  as mentioned in \cite{cf}. For the specific cases of spin-2 and spin-3, we work out here such connection explicitly and we go even beyond the Maxwell-like sub case. In section 2 we recover the results of \cite{blas} and \cite{rr} for the gauge invariant (saturated) two point amplitude of the generalized spin-2 TDiff model. In section 3 we define  a generalized spin-3 TDiff model and obtain its two point gauge invariant amplitude  using the rank-3 projection operators given in the appendix.

Since the doublet action is invariant under unconstrained Diff symmetry, from the perspective of TDiff models, we develop a method in section 4 to lift constrained symmetries into unconstrained ones  by introducing appropriate
Stueckelberg-like fields\footnote{See the review work\cite{sagnotti_notes} and \cite{cf} for the introduction of Stueckelberg fields in HS theories along the same goal.}, thereby connecting the one-field (singlet) generalized TDiff models with the string theory doublet action in a direct manner. It turns out that a perfect match only occurs apparently for specific points in the parameters space of the TDiff models .

Finally, in section 5, motivated by the apparent mismatch of section 4, we show that there is a non trivial equivalence among the physical spin-2 and spin-3 TDiff models such that, for each spin, it is always possible to start from one point in physical subspace  of  the parameters of the TDiff model and after a local field redefinition and sources redefinitions, to arrive at any other point of this subspace. In particular, any model is equivalent to its simplest version, called Maxwell-like model.

In section 6 we draw our conclusions and perspectives for future work in order to generalize the equivalence we have found for higher spins, curved spaces and mixed symmetry tensors.

\section{$s=2$ TDiff model - A review}\label{sec:s=2rev}

Before we go to the spin-3 case we review the spin-2 TDiff model since there are several parallel aspects which inspire us to run after a generalization of our results for arbitrary spin-$s$.

The free spin-2 TDiff model is
written in terms of a symmetric rank-2 field $h_{\mu\nu}=h_{\nu\mu}$.
Following the notation\footnote{ We use  the Minkowski metric $\eta_{\mu\nu}=diag(-1,1,...,1)$ in $D$ dimensions and our rank-2 and rank-3 index symmetrization have no weights, e.g., $T_{(\alpha\beta)} = T_{\alpha\beta} + T_{\beta\alpha}$.} of \cite{rr} we have in $D \ge 3$ dimensions,
\be S(a,b) = \int d^Dx\, \left\lbrack\frac{1}{2} h^{\alpha\beta}\, \Box \,  h_{\alpha\beta}+(\del^\mu h_{\mu\beta})^2+ a\, h\, \del^{\mu}\del^{\nu} h_{\mu\nu}-\frac{b}{2} h\,  \Box h \, + h_{\mu\nu} T^{\mu\nu}\right\rbrack .
\label{tds2}
\ee
\no The first two terms are required for massless spin-2 propagation without
spin-1 ghosts \cite{blas}. Neglecting surface terms, for any $a,b\in\mathbb{R}$, the action (\ref{tds2}) is invariant under TDiff gauge transformations:
\be 
\delta_{T\,} h_{\mu\nu} = \del_\mu \Lambda^{T}_{\nu} +  \del_\nu \Lambda^{T}_{\mu}\quad;\quad \del^\mu \Lambda^T_\mu=0 \, ,
\label{tdiffs}
\ee
if the source satisfies:

\be \p^{\mu}T_{\mu\nu} = \p_{\nu}J  \quad , \label{source1}\ee

\no where  $J$ is an arbitrary function. So the original source is in general not conserved. 

The Lagrangian corresponding to (\ref{tds2}) can be written in the form

\be {\cal L}(a,b) = \frac 12 \, h_{\mu\nu}G^{\mu\nu\alpha\beta}(\p)h_{\alpha\beta} + \, h_{\mu\nu} \, T^{\mu\nu} \quad , \label{tds2g} \ee

\no where the second order differential operator $G(\p)$ is read off from (\ref{tds2}). The gauge invariant two point amplitude in momentum space is given by the propagator saturated with the sources,  in our notation and suppressing indices,

\be \mathcal{A}_2 (k) = -\frac{i}{2} T^* (k)G^{-1}(\p \to i\, k) T(k) \quad.   \label{a2k} \ee 

  We can check the particle content of (\ref{tds2}) by analysing the analytic structure of $ \mathcal{A}_2 (k)$. Following \cite{rr}, see also \cite{blas}, in order to obtain $G^{-1}(k)$,  we add a gauge fixing term in the form of a square of  a transverse vector. Explicitly, in momentum space we end up with
  
  \be 
\mathcal{A}_2 (k) = \, \frac{i}{k^2}\, \left[\tilde{T}^{*\mu\nu}\tilde{T}_{\mu\nu}-\frac{|\tilde{T}|^2}{D-2}+\frac{1}{(D-2)f_{D}}\,|\tilde{J}|^2 \right] \label{ampldiag}.
\ee

\no where the conserved tensor $\tilde{T}_{\mu\nu} $  and $\tilde{J}$ are given  by:

\be
\tilde{T}_{\mu\nu}=T_{\mu\nu}-\eta_{\mu\nu}J \quad ; \quad 
\tilde{J}= (1-a)\, T+ (aD-2)J; \label{tildetj}
\ee

\no with $\del^\mu\tilde{T}_{\mu\nu}=0$ and  $T=\eta^{\mu\nu}T_{\mu\nu}$. The  constant $f_D$  \cite{blas} is given by

\be f_D \equiv f_D (a,b) = (D-2)(a^2-b) + (a-1)^2. \quad \label{fdab} \ee

\no The first two terms of (\ref{ampldiag}) represent a physical massless spin-2 particle. They coincide with the amplitude of the linearized Einstein-Hilbert theory. The last term represents a physical (unphysical) massless spin-0  particle if $f_D >0 $ ($f_D < 0$). Such content can be confirmed by a purely Lagrangian analysis, see \cite{blas}, or by the usual Hamiltonian Dirac-Bergmann algorithm for constrained systems as carried out in \cite{masterrr}. In \cite{rr} one has shown that the spin-0 term leads to an additive contribution to the deviation angle in the gravitational lens effect, as such, it could be used to explain an eventual experimental deviation above the angle predicted by general relativity  but not below it. 

The singularity in the amplitude at $f_D=0$ points to the appearance of a larger symmetry. In particular, at $(a,b)=(1,1)$ we have the linearized Einstein-Hilbert (EH) theory invariant under unconstrained Diff, i.e., 
$ \delta h_{\mu\nu} = \del_\mu \Lambda_{\nu} +  \del_\nu \Lambda_{\mu}$ while at $(a,b)=(a_D,b_D)\equiv (2/D,(D+2)/D^2)$ we have the spin-2 WTDiff model \cite{blas} where $ \delta_{WT} h_{\mu\nu} = \del_\mu \Lambda^T_{\nu} +  \del_\nu \Lambda^T_{\mu} + \eta_{\mu\nu} \, \lambda $. The other cases, as we explain later, are trivial shifts of the EH case. Whenever we have $f_D=0$ we have the propagation of only massless spin-2 particles.

\section{$s=3$ TDiff model}

In order to be as close as possible to the spin-2 case, regarding the notation, we start this subsection with the most general second order spin-3 massless model free of rank-2 ghosts. In terms of a fully symmetric rank-3 tensor $h_{\mu\nu\rho}=h_{(\mu\nu\rho)}$ we have

\be S(a,b,c) = \int d^Dx\, \left\lbrack \frac 12 h_{\mu\nu\rho}\, \Box \, h^{\mu\nu\rho} + \frac 32 (\p^{\mu}h_{\mu\nu\rho})^2 + 3\, a \, \p^{\mu}\p^{\nu} h_{\mu\nu\rho} \, h^{\rho}- \frac 32\, b \, h_{\mu}\, \Box \, h^{\mu} + \frac 34\, c \, (\p^{\mu} \, h_{\mu})^2 \right\rbrack. \\ \label{sabc} \ee

\no The trace is denoted by the vector $h_{\mu} = \eta^{\nu\rho} h_{\mu\nu\rho}$, while $(a,b,c)$ are so far arbitrary real constants. The first two terms in (\ref{sabc}) are required for a massless spin-3 particle without spin-2 ghosts. The action $S(a,b,c)$  is invariant under traceless and transverse higher spin diffeomorphisms (TrTDiff henceforth) depicted by a bar and a superscript ``T'' respectively, i.e., 

\be \dlb h_{\mu\nu\rho} = \p_{(\mu}\bar{\Lambda}^T_{\nu\rho)} \qquad ; \qquad  \p^{\nu}\bar{\Lambda}^T_{\nu\rho} = 0 = \eta^{\nu\rho}\bar{\Lambda}^T_{\nu\rho} \, . \label{difftt} \ee 

\no The number of independent gauge parameters of the TrTDiff symmetry is not enough to get rid of all nonphysical modes, the gauge symmetry must be enlarged. There are by now in the literature, to the best we know, three unitary models containing physical massless spin-3 particles which are subcases of  (\ref{sabc}) with a larger symmetry. They have first appeared in \cite{fronsdal}, \cite{sv} and \cite{cf}. Those models and their gauge symmetries are given respectively by:

\begin{table}[h!]
	\centering
	\renewcommand{\arraystretch}{1.2} 
	\begin{tabular}{|p{5cm}|p{6cm}|}
		\hline
		\textbf{Action} & \textbf{ Type of Symmetry - $\delta h_{\mu\nu\rho}$} \\ 
		\hline\hline 
		\( S_F  \equiv S(1,1,1)\) & \( \partial_{(\mu} \bar{\Lambda}_{\nu\rho)} \), \quad\quad\quad\quad\quad\,\, \((\mathrm{TrDiff}) \) \\ 
		\hline
		\( S_{SV} \equiv S(a_{D+2},b_{D+2},c_{D+2}) \) & \( \partial_{(\mu} \bar{\Lambda}^T_{\nu\rho)} + \eta_{(\mu\nu}\lambda_{\rho)} \),\quad\quad \((\mathrm{WTrTDiff} )\) \\ 
		\hline
		\( S_{Max} \equiv S(0,0,0) \) & \( \partial_{(\mu} \Lambda^T_{\nu\rho)} \),\quad\quad\quad\quad\quad\,\,\, \(( \mathrm{TDiff}) \) \\ 
		\hline
	\end{tabular}
	\caption{Gauge transformations, actions, and their corresponding symmetries.}
	\label{tab:gauge_symmetries_combined}
\end{table}

\no In the table-1, see \cite{sv},  we have: 
 
 \be (a_{D+2},b_{D+2},c_{D+2}) = \left(\frac 2{D+2},\,\, \frac{D+4}{(D+2)^2},\,\, \frac{2(2-D)}{(D+2)^2}\right) \quad . \label{abcstar} \ee 
 
In the cases of the Fronsdal ($S_F$) and the Skvortsov-Vasiliev ($S_{SV}$) models, the number of independent gauge parameters is the same\footnote{The Skvortsov-Vasiliev (SV) model can be alternatively formulated in terms of a traceless rank-3 symmetric  field where one is left only with the TrTDiff  symmetry. One has used the vector Weyl symmetry to make $h_{\mu}=0$. Since we focus in the connection with the spectrum of the tensionless limit of string field theory where no such restrictions on the rank-$s$ field show up,  we work only with unrestricted fields.}, $D(D+1)/2-1$, just like their spectrum which consists of a physical massless spin-3 particle. However, in the Maxwell-like ($S_{Max}$) model the TDiff symmetry has less parameters, $D(D-1)/2$. Consequently, besides the physical spin-3 particle, we also have a physical massless spin-1 particle.

 Next we investigate the most general spin-3 case, i.e., $S=S(a,b,c)$. First of all we introduce a generalization of the TDiff transformation\footnote{Such generalization only exists for $s\ge 3$ since for $s=2$ we have $\p_{\mu}\Lambda_{\nu}^T + \p_{\nu}\Lambda_{\mu}^T + k \, \eta_{\mu\nu} \p^{\rho}\Lambda_{\rho}^T = \p_{\mu}\Lambda_{\nu}^T + \p_{\nu}\Lambda_{\mu}^T$.}  via a real constant $k$, i.e., 
 
 \be \delta \, \hmn = \p_{(\mu}\Lambda^T_{\nu\rho)} + k\, \eta_{(\nu\rho}\p_{\mu)} \, \Lambda^T = \p_{(\mu}[\Lambda^T_{\nu\rho)} + k \, \eta_{\mu\nu)}\Lambda^T] \equiv \p_{(\mu}\tilde{\Lambda}_{\nu\rho)} \, , \label{gtdiff} \ee
 
 \no where $\Lambda^T = \eta^{\mu\nu}\Lambda_{\mu\nu}^T$. Accordingly we have,
 
 \be \delta\, S(a,b,c) = 3 \int d^Dx \, \Lambda^T \Big\{ \p^{\mu}\p^{\nu}\p^{\rho} \hmn \Big[ k(2-a(D+2))- a\Big] + \Box \,\p^{\mu}h_{\mu}\Big[\Big(b + \frac c2\Big)(k(D+2) +1) -3\, a\, k\Big]\Big\} \, . \label{deltas} \ee
 
 \no  Thus, if $a=a_{D+2}=2/(D+2)$ there will be no  TDiff symmetry unless we also require a traceless parameter $\Lambda^T=0$. In the next subsection we assume $a\ne 2/(D+2)$ which leads to TDiff models closer to the Fronsdal theory. Later on we will investigate the special case $a=a_{D+2}=2/(D+2)$, closer to the SV theory.

\subsection{Case I: $a \ne 2/(D+2)$}

 Requiring  invariance of the general action i.e., $\delta\, S(a,b,c)=0$, we have from (\ref{deltas}):
  
  \be k = \frac a{2-a(D+2)} \qquad ; \qquad c= 3\, a^2 - 2 \, b \, \equiv c_S . \label{cstar} \ee 
 
 Remarkably, the special case $c = c_s$ also holds for all spin-3 models
mentioned earlier in Table-1.
 Regarding the gauge parameter  $\tilde{\Lambda}_{\mu\nu}$ defined in (\ref{gtdiff}), if $a\ne 1$, it satisfies the vector constraint:

\be \p^{\nu}\tilde{\Lambda}_{\nu\rho} = \frac k{1+k\, D} \,\p_{\rho} \tilde{\Lambda} = \frac a{2(1-a)}\, \p_{\rho} \tilde{\Lambda} \, . \label{vector} \ee

 \no If $a= 1$ the  parameter must be traceless ($\eta^{\nu\rho}\tilde{\Lambda}_{\nu\rho} = \tilde{\Lambda} =0$) and the transverse part of its divergence must   vanish: $\left( \p^{\nu}\tilde{\Lambda}_{\nu\rho}\right)^T =0$. So in total we still have  $D$ constraints as in the $a\ne 1$ case. This shows that (\ref{gtdiff}) is indeed a natural analogue for $a\ne 0$ of the usual TDiff symmetry $ \delta \, \hmn = \p_{(\mu}\Lambda^T_{\nu\rho)}$ which only occurs at $a=0$.
 
 In order to check the particle content of (\ref{sabc}) with $c= c_S $ we introduce a totally symmetric source as in the $s=2$ case. Namely, we define
 
 \be S(a,b,c_S,T_{\mu\nu\rho},J_{\mu}) \equiv S(a,b,c_S) + \int d^D x \,\, \hmn T^{\mu\nu\rho}  \quad . \label{sts} \ee 
 
 \no The gauge invariance of the source term, after integration by parts, imply the following  constraints:
 
 \bea \p_{\mu}\, T^{\mu\nu\rho} &=& \p^{\nu}\, J^{\rho} + \p^{\rho}\, J^{\nu} 
 - \frac{2\, k\, \eta^{\nu\rho} }{(1+k\, D)} \, (\p^{\mu}J_{\mu}) \quad ; \quad a\ne 1 \,\, \mathrm{and} \, \, (a,b)\ne (a_{D+2},b_{D+2}) \, , \label{can1} \\
 \p_{\mu}\, T^{\mu\nu\rho} &=& \p^{\nu}\, J_T^{\rho} + \p^{\rho}\, J_T^{\nu} 
 + \eta^{\nu\rho} \, J \quad\quad\quad\quad\quad\quad ; \quad a= 1 \,\, \mathrm{and} \, \, b\ne 1 \, , \label{ca1} \eea
 
\no where $\p^{\mu}J_{\mu}^T =0$ while $J_{\mu}$ and $J$ are arbitrary functions. In particular, there is no connection between $J_{\mu}^T$ and $J_{\mu}$. The restrictions on the far right of (\ref{can1}) and (\ref{ca1}) are necessary to avoid the SV and Fronsdal cases respectively, where we have more symmetry and the constraints are more restrictive, namely, 

\bea \p_{\mu}T^{\mu\nu\rho} &=& \eta^{\nu\rho} \, J  \qquad \quad\quad\quad\quad\quad\quad\quad\quad\quad\quad\quad ({\rm Fronsdal}) \label{cfron} \\
\p_{\mu}T^{\mu\nu\rho} &=& \p^{\nu}\, J_T^{\rho} + \p^{\rho}\, J_T^{\nu} 
 \quad ; \quad \eta_{\mu\nu}T^{\mu\nu\rho} = 0  \qquad\,\, ({\rm SV}),  \label{csv} \eea
 
 \no where $J_{\mu}^T$ and  $J$ are arbitrary except for $\p^{\mu}J_{\mu}^T=0$  . After adding a gauge fixing term ${\cal L}_{GF} = \lambda_0 \, (f_{\mu\nu}^T)^2/2 $ to (\ref{sts}), 
 where we have defined the symmetric transverse tensor:
 
 \be f_{\nu\rho}^T = \Box \, \p^{\mu}\hmn - \p_{\nu}(\p^{\alpha}\p^{\beta}h_{\alpha\beta\rho}) - \p_{\rho}(\p^{\alpha}\p^{\beta}h_{\alpha\beta\nu}) + \eta_{\nu\rho}\, \p^{\alpha}\p^{\beta}\p^{\gamma}h_{\alpha\beta\gamma} \quad ,
 \label{fmn} \ee
 
 \no we are able to obtain the rank-3 tensor propagator. Saturating with sources according to (\ref{a2k}) the dependence on $\lambda_0$ disappears and we obtain the two point amplitude in momentum space associated with $S(a,b,c_S)$. We have carried out this   procedure separately for $a\ne 1$ and $a=1$. In the first case, assuming also $(a,b)\ne (a_{D+2},b_{D+2})$,  we have:
 
  \be 
\mathcal{A}_2(k)= \, \frac{2i}{k^2}\, \left[\tilde{T}^*_{\mu\nu\rho}\tilde{T}^{\mu\nu\rho}-\frac 3D \, |\tilde{T}_{\mu}|^2 \, + \, \frac{3}{D\, f_{D+2}}\,|\tilde{J}^T_{\mu}|^2 \right] \label{a2ks3};
\ee
 
\no where:

\bea \tilde{T}^{\mu\nu\rho} &\equiv &  T^{\mu\nu\rho} \, - \eta^{(\mu\nu}\, J^{\rho)} \quad , \label{ttil} \\
\tilde{J}^T_{\mu} &\equiv & (2- a(D+2))  J_{\mu} \, + \, (a-1)\, T_{\mu} \quad , \label{jtil} \\
f_{D+2} (a,b) &\equiv & D(a^2-b)+(a-1)^2 \quad . \label{fb} \eea 

\no The first two terms in (\ref{a2ks3}) correspond precisely to the result one finds in the spin-3 Fronsdal theory. Notice that from (\ref{can1}) and (\ref{cstar}) one can show that $\p_{\mu}\tilde{T}^{\mu\nu\rho} = \eta^{\nu\rho} \, (\p^{\mu}J_{\mu})/(a-1)$ which is a constraint of the Fronsdal type, see (\ref{cfron}). Likewise, (\ref{can1}), (\ref{cstar}) and (\ref{jtil}) lead to   $\p^{\mu} \tilde{J}^T_{\mu}=0$. Moreover, it turns out that in the second case ($a=1$), using the constraint (\ref{ca1}), we end up basically with the same expression (\ref{a2ks3}) with the replacements $\tilde{T}^{\mu\nu\rho} \to  T^{\mu\nu\rho} \, - \eta^{(\mu\nu}\, J_T^{\rho)}$ and $(f_{D+2},\tilde{J}_{\mu}^T) \to (D\,(1-b),J_{\mu}^T)$. In summary, regarding the particle content, the expression (\ref{a2ks3}) holds in principle
 for any $a\ne 2/(D+2)$
and we have always a physical massless spin-3 particle plus a physical (non physical) massless spin-1 particle if $f_{D+2}>0$ ( $f_{D+2}<0$ ). Notice that this is in agreement with the interpretation of the spin-3 version of the Maxwell-like models of \cite{cf} as describing physical spin-3 and spin-1 massless particles in the case of rank-3 symmetric tensors, since $f_{D+2}(0,0)=1 >0$. As in the spin-2 case, if $f_{D+2}=0$ there is an enlargement of the symmetry and we only have massless spin-3 particles.

\subsection{Case II: $a = 2/(D+2)$}

As we have mentioned  after (\ref{deltas}), if $a=a_{D+2}= 2/(D+2)$ in order to have a symmetry we have to  assume further that the transverse rank-2 gauge parameter is  traceless (TrTDiff). It turns out, as the reader can check, that if $c=c_S=3\, (a_{D+2})^2 - 2\, b$ there will be also a longitudinal Weyl  symmetry $\delta\, \hmn= \eta_{(\mu\nu}\p_{\rho)}\lambda $ that is enlarged to full Weyl $\delta\, \hmn= \eta_{(\mu\nu}\lambda_{\rho)} $ at the SV point $b=b_{D+2}$. In this subsection we assume $(a,b,c)=(a_{D+2},b, c_S)$ with $b\ne b_{D+2}$. Consequently, the action
$S(a_{D+2},b,c_S)$ is invariant under the following transformation 
for otherwise arbitrary values of $b$,

\be \delta\, \hmn = \p_{(\mu}\bar{\Lambda}^T_{\nu\rho)} + \, \eta_{(\mu\nu}\p_{\rho)}\lambda \equiv \p_{(\mu}\Lambda^*_{\nu\rho)} \quad . \label{lstar} \ee

\no The traceless piece of the new gauge parameter must be transverse, i.e., similarly to (\ref{vector}), it satisfies a vector constraint: 

\be \p^{\nu}\Lambda^*_{\nu\rho} = \frac 1D \p_{\rho}\Lambda^* \quad . \label{vector2} \ee

\no Notice that (\ref{vector2}) does not fit in the form (\ref{vector}), there would be no solution for $k$. Henceforth we call T$^*$Diff the diffeomorphisms (\ref{lstar}) satisfying (\ref{vector2}). The invariance of (\ref{sts}) under  (\ref{lstar}), using (\ref{vector2}) at $a=a_{D+2}$ requires the constraint


\be \p_{\mu}T^{\mu\nu\rho} = \p^{\nu}J^{\rho} + \p^{\rho}J^{\nu} -\frac 2D\, \eta^{\nu\rho} \p_{\mu}J^{\mu} \quad , \label{ctsdiff} \ee 

\no where $J_{\mu}$ is arbitrary. Since the number of independent gauge parameters of T$^*$Diff is the same one of the generalized TDiff we can still use the same gauge fixing term based on (\ref{fmn}). After saturating with sources the gauge fixing constant $\lambda_0$ disappears again and we recover (\ref{a2ks3}) at $a=a_{D+2}$ where now $f_{D+2}(a_{D+2},b)=D(b_{D+2}-b) $ and $\tilde{J}_{\mu}^T=T_{\mu}.$ From the constraint (\ref{ctsdiff}) we have $\p^{\mu} \tilde{J}^T_{\mu}=\p^{\mu}T_{\mu} =0$ and $\p_{\mu} \tilde{T}^{\mu\nu\rho} = -(D+2)\eta^{\nu\rho}\p\cdot J/D $ which, as in the $a\ne 2/(D+2)$ case, allows us to conclude that $S(a_{D+2},b,c_S)$ contains a physical spin-3 particle plus a physical (non physical) massless spin-1 particle for $b<b_{D+2}$ ($b>b_{D+2}$). The singularity at $b=b_{D+2}$ indicates the appearance of a larger symmetry, in this case the longitudinal Weyl symmetry in (\ref{lstar}) becomes the full Weyl symmetry.

\section{Doublet action: TDiff $\to$ Diff and T$^*$Diff $\to$ Diff } 

It is known that consistent propagations of strings 
require specific (critical) spacetime dimensions like $D=26$ ($D=10$) for the bosonic (super) string. By choosing a critical dimension we cancel the central charge of the original Virasoro algebra. It is possible however, see \cite{ouvry,beng,ht}, 
 to redefine the Virasoro generators in the tensionless limit 
($1/\alpha' \to 0$) such that their new algebra has no central charge and no critical dimension is required. Using the corresponding BRST charge one can construct a string field theory which in the open bosonic string case, restricted to symmetric fields, leads to a quadratic second order action written in terms of a triplet of totally symmetric  fields ($\vf,C,D$) of rank $(s,s-1,s-2)$ respectively, see \cite{fscqg,st}.   The corresponding theory describes free massless particles of  spin  $s, s-2, s-4,\cdots, 0$ or 1 according to even or odd integer s, in arbitrary dimensions. After algebraically eliminating the field $C$ one obtains a doublet action in terms of ($\vf,D$). Here, we change the notation 
$D_{\mu_1\cdots \mu_{s-2}} \to  \chi_{\mu_1\cdots \mu_{s-2}} $ and the doublet action of \cite{st} becomes\footnote{We use the compact notation : $(\partial \cdot)^k \vf = \partial^{\mu_1} \cdots \partial^{\mu_k} \vf_{\mu_1 \dots \mu_k \mu_{k+1} \cdots \mu_s}$.}

\be S_{ST} = \int d^Dx \Big\{ \frac 12 \vf \, \Box \, \vf + \frac s2 (\p \cdot \vf)^2 +
 s(s-1)\left[\chi\, \p\cdot \p\cdot \, \vf - \chi \, \Box \, \chi \right] + 3\, \binom{s}{3} (\p\cdot \chi)^2 \Big\} \, . \label{st} 
\ee

Unlike the Fronsdal theory, in (\ref{st}) there is no
double traceless condition on the fields; both are unconstrained. Moreover there is no constraint in the gauge symmetry parameter either, the action (\ref{st}) is invariant under unconstrained diffeomorphism:

\be \delta\, \vf_{\mu_1 \cdots \mu_s} = \p_{(\mu_1}\Lambda_{\mu_2\cdots \mu_s)} \qquad ; \qquad \delta\, \chi_{\mu_1 \cdots \mu_{s-2}}  = \p^{\rho}\Lambda_{\rho\mu_1\cdots \mu_{s-2}} \, . \label{diffs} \ee

 In order to compare both $s=2$ and $s=3$ TDiff models (\ref{tds2}) and 
 (\ref{sabc}) respectively  with the corresponding cases in  (\ref{st})
 we must first lift the transverse condition  on the gauge parameter turning TDiff into Diff. We will carry this out in both cases $s=2$ and $s=3$.
We start with the $s=2$ case but basically the same procedure will be performed for $s=3$.

\subsection{ From TDiff $\to$ Diff  ($s=2$) }

We first rewrite the transverse parameter in terms of an arbitrary one via a non local projection operator:
 
 \be \Lambda_{\mu}^T = \Lambda_{\mu} - \frac{\p_{\mu}(\p^{\rho}\Lambda_{\rho})}{\Box} \quad , \label{lambdat} \ee
 
 \no such that the field in (\ref{tds2}) transforms as
 
\be  \delta_{T\,} h_{\mu\nu} = \del_\mu \Lambda^{T}_{\nu} +  \del_\nu \Lambda^{T}_{\mu} = \del_\mu \Lambda_{\nu} +  \del_\nu \Lambda_{\mu} -2 \, \frac{\p_{\mu}\p_{\nu}(\p\cdot\Lambda)}{\Box} \quad . \label{tdh2} \ee

\no The previous formula suggests the introduction  of a compensating scalar field via a non local  field redefinition :

\be h_{\mu\nu}  = \vf_{\mu\nu} -2 \,\frac{\p_{\mu}\p_{\nu}\chi}{\Box} \quad , \label{fredefs2} \ee

\no such that the new fields transform locally under unconstrained diffs, 

\be \delta \, \vf_{\mu\nu} =  \del_\mu \Lambda_{\nu} +  \del_\nu \Lambda_{\mu} \qquad ; \qquad \delta \, \chi = \p^{\mu}\Lambda_{\mu} \quad , \label{diffs2} \ee

\no which makes us to expect the cancellation of the non local terms in the action (\ref{tds2}). Indeed, after (\ref{fredefs2}), its Lagrangian density is local, including the source term, 

\bea {\cal L}^{s=2}[a,b,\vf,\chi] &=&  \frac{1}{2} \vf^{\alpha\beta}\, \Box \,  \vf_{\alpha\beta}+(\del^\mu \vf_{\mu\beta})^2+a\, \vf \,\del^\mu\del^\nu \vf_{\mu\nu}-\frac{b}{2}\vf \, \Box\,\vf \nn\\ &+& 2(b-a)\chi \, \Box\, \vf + 2(1-a)\,\chi \p\cdot\p\cdot \vf + 2\, C\, \chi \, \Box \, \chi + \vf_{\mu\nu}T^{\mu\nu} + \chi\, J\, , \label{ldiffs2} \eea

\no where
$C\equiv  2\, a - b -1 $ and we have used the source constraint (\ref{source1}). The action corresponding to (\ref{ldiffs2}) is now invariant under the unconstrained diffeomorphism in (\ref{diffs2}). 

 Regarding the particle content of (\ref{ldiffs2}), after the invertible field redefinition, see \cite{rr},

\be \left(\vf_{\mu\nu}\,;\,\chi \right)\to \left(H_{\mu\nu}+ \frac{(1-a)}{D-2}\eta_{\mu\nu}\Phi\, ;\, - H + \frac{(aD-2)}{D-2}\,\Phi \right) \quad , \label{fredfdiag2} \ee

\no we have a diagonal theory consisting of a linearized Einstein-Hilbert theory (healthy spin-2) and an additional  spin-0 particle, 

\be {\cal L}^{s=2}[a,b,H,\Phi] = {\cal L}_{LEH}(H) + \frac{f_D(a,b)}{D-2} \, \Phi \, \Box \, \Phi + H_{\mu\nu}\tilde{T}^{\mu\nu} + \Phi\, \tilde{J}. \label{ldiags2} \ee 

\no The functional integral over $(H_{\mu\nu},\Phi)$ will exactly reproduce 
 (\ref{ampldiag}), showing that the particle content of (\ref{tds2}) has not been changed by the nonlocal introduction of $\chi$ via (\ref{fredefs2}).
 
  For future comparison with 
the doublet action (\ref{st}) coming from strings at $s=2$, we rewrite (\ref{ldiffs2}) in the equivalent form\footnote{Note that $\varphi$ in the second line of (\ref{stgs2}) stands for the trace of the rank-2 field $\varphi=\eta^{\mu\nu}\varphi_{\mu\nu}$ except in $\p\cdot\p\cdot \varphi$.}:

\bea {\cal L}^{s=2}[a,b,\vf,\chi] &=&  \frac 12 \vf \, \Box \, \vf + (\p \cdot \vf)^2 +
 2\,\chi\, (\p\cdot \p\cdot \, \vf) - 2\, \chi \, \Box \, \chi \nn\\
 &+& \frac{(2\, a-b)}2(\vf-2\, \chi)\, \Box \, (\vf-2\, \chi) + a(\vf-2\, \chi)(\p\cdot\p\cdot \vf- \Box \vf) \quad . \label{stgs2} \eea

\subsection{ From TDiff $\to$ Diff  ($s=3$) }

 Now we move to the $s=3$ case. First we look at the  $\bm{a \ne 2/(D+2)}$ case. The Diff transformation 
 $\delta\, h_{\mu\nu\rho} = \p_{(\mu}\tilde{\Lambda}_{\nu\rho)}$ which leaves (\ref{sabc}) invariant, at $c=c_S$, is such that the vector constraint (\ref{vector}) must hold. It is a consequence of $\tilde{\Lambda}_{\nu\rho}=  \Lambda_{\nu\rho}^T + \, k\, \eta_{\nu\rho}\, \Lambda^T$. Following the steps of the $s=2$ case we write down $ \Lambda_{\nu\rho}^T$ in terms of an arbitrary rank-2 symmetric field via a projection operator. Consequently, we have, 
\be \tilde{\Lambda}_{\nu\rho} = \Lambda_{\nu\rho}- \frac{\p_{(\nu}\p^{\mu}\Lambda_{\mu\rho)}}{\Box}  + \, \eta_{\nu\rho}\, \left(k\, \Lambda + g\,\frac{\p\cdot\p\cdot\Lambda}{\Box}\right) \, , \label{ltilf} \ee

\no where $k$ is given in (\ref{cstar}) and $ g=1+k\, (D-2)$. The reader can check that (\ref{ltilf}) satisfies (\ref{vector}). 

Then it follows that: 

 \be \delta \, \hmn = \p_{(\mu}\tilde{\Lambda}_{\nu\rho)} = \p_{(\mu}\Lambda_{\nu\rho)} - 2\, \frac{\p_{(\mu}\p_{\nu}\lambda_{\rho)}}{\Box} + \eta_{(\mu\nu}\, \p_{\rho)}\left( k\, \Lambda + g\, \frac{\p\cdot\lambda}{\Box}\right) \, , \label{dell} \ee
 
 \no where we have defined $\lambda_{\mu} \equiv \p^{\rho}\, \Lambda_{\rho\mu} $. If we introduce a rank-3 ($\vf$) and a rank-1 ($\chi$) field which transform according to the spin-3 version of (\ref{diffs}), i.e.:
 
 \be \delta\, \vf_{\mu\nu\rho} = \p_{(\mu}\Lambda_{\nu\rho)} \qquad ; \qquad \delta\, \chi_{\mu} = \lambda_{\mu} = \p^{\rho}\, \Lambda_{\rho\mu} \, , \label{diffs3} \ee 
 
 \no we can rewrite the right side of (\ref{dell}) in terms of $\delta \vf_{\mu\nu\rho}$ and $\delta\chi_{\mu}$. The resulting expression is in agreement with the non local field redefinition:
 
 \be  \hmn = \vf_{\mu\nu\rho}  + \eta_{(\mu\nu} V_{\rho)}+2\, \frac{\p_{(\mu}\p_{\nu}\chi_{\rho)}}{\Box}  \, , \label{fredfs3} \ee
 where
\be  V_{\mu} = k\, (\vf_{\mu} + 2\, \chi_{\mu}) - g\, \frac{\p_{\mu}(\p\cdot\chi)}{\Box} \, . \label{V} \ee

The substitution of (\ref{fredfs3}) in (\ref{sabc}), leads to the following local theory:

\bea {\cal L}_{s=3}[a,b,\vf,\chi] &=& \frac 12 \vf_{\mu\nu\rho}\, \Box \, \vf^{\mu\nu\rho} + \frac 32 (\p^{\mu}\vf_{\mu\nu\rho})^2 - \frac 32 f\, \vf^{\mu}\, \Box \, \vf_{\mu} - \frac 32 f\, (\p\cdot \vf)^2 \nn\\
&+& 6\, \chi\, (\p\cdot\p\cdot \vf) +6 \, f \, \chi^{\mu}\, \Box \, \vf_{\mu} + 6 \, f\, (\p\cdot \chi)(\p\cdot \vf) \nn \\
&-& 6 \, (1+f)\chi^{\mu}\, \Box \, \chi_{\mu} -3 \,(2\, f-1)(\p\cdot \chi)^2 \, , \label{ls3ab} \eea

\no where 

\be f = \frac{a^2\, D -4\, a + 4\, b}{[a(D+2)-2]^2} \quad . \label{g} \ee

\no  After the invertible field redefinition:

\be (\vf_{\mu\nu\rho}\,,\,\chi_{\mu})= \left( H_{\mu\nu\rho} - \eta_{(\mu\nu}A_{\rho)} \, ,\, A_{\mu} - \frac{H_{\mu}}{2} \right) \, , \label{fredefs3diag} \ee

\no the Lagrangian (\ref{ls3ab}) acquires a diagonal form as a linear combination of the spin-3 Fronsdal theory and the Maxwell theory, 

\be {\cal L}_{s=3}[a,b] = {\cal L}_F(1,1,1,H) -\,  \frac{3D\, f_{D+2}}{[a(D+2)-2]^2} \, F_{\mu\nu}^2(A) \quad , \label{f+m} \ee

\no  where ${\cal L}_F(1,1,1,H)$ is the Fronsdal Lagrangian corresponding to  (\ref{sabc}) with $(a,b,c)=(1,1,1)$. So besides the physical massless spin-3 particle, if $f_{D+2} >0$ ($f_{D+2} <0$)
we have a physical (non physical) massless spin-1 particle which disappears at $f_{D+2}=0$ where TDiff increases to TrDiff. Notice that this is in full agreement with the two point amplitude (\ref{a2ks3}). So once again the non local introduction of Stueckelberg-like field ($\chi_{\mu}$) did not change the particle content of the model.

For sake of comparison with the doublet action (\ref{st}) we rewrite (\ref{ls3ab}) in the equivalent form:

\bea {\cal L}_{s=3}[a,b,\vf,\chi] &=& \frac 12 \vf_{\mu\nu\rho}\, \Box \, \vf^{\mu\nu\rho} + \frac 32 (\p^{\mu}\vf_{\mu\nu\rho})^2
+ 6\, \chi\, (\p\cdot\p\cdot \vf) - 6 \, \chi^{\mu}\, \Box \, \chi_{\mu} +3 \,(\p\cdot \chi)^2 \nn\\ &+& \frac{3\, f}2 F_{\mu\nu}^2(\vf-2\, \chi)\,\,.  \label{stgs3} \eea

\subsection{ From T$^*$Diff $\to$ Diff  ($s=3$) }

Now we look at the $\bm{a =a_{D+2}= 2/(D+2)}$ case. The T$^*$Diff transformation 
 $\delta\, h_{\mu\nu\rho} = \p_{(\mu}\Lambda^*_{\nu\rho)}$ now must satisfy the constraint (\ref{vector2}). Similarly to the previous cases
 the reader can check that 
 
\be \Lambda^*_{\nu\rho} = \Lambda_{\nu\rho}- \frac{\p_{(\nu}\p^{\mu}\Lambda_{\mu\rho)}}{\Box}  + \, \frac{\p_{\nu}\p_{\rho}}{(D-1)\, \Box} \left\lbrack \Lambda + (D-2)\frac{\p\cdot\p\cdot\Lambda}{\Box} \right\rbrack \, , \label{lstarnl} \ee

\no indeed satisfies (\ref{vector2}). From (\ref{lstarnl}) we can write:

 \be \p_{(\mu}\Lambda^*_{\nu\rho)} = \p_{(\mu}\Lambda_{\nu\rho)} + 2\, \frac{\p_{(\mu}\p_{\nu}U_{\rho)}}{\Box} -\frac {3(D-2)}{D-1}\, \frac{\p_{\mu}\p_{\nu}\p_{\rho}(\p\cdot \, U)}{\Box^2}  \, , \label{dellstar} \ee
 
 \no where we have defined $U_{\mu} \equiv - (\p\cdot\Lambda)_{\mu} + \p_{\mu}\Lambda/D$  . Therefore, if we introduce, as before, a rank-3 ($\vf$) and a rank-1 ($\tchi$) field which transform according to unconstrained diffs as :
 
 \be \delta\, \vf_{\mu\nu\rho} = \p_{(\mu}\Lambda_{\nu\rho)} \qquad ; \qquad \delta\, \tchi_{\mu} =  -\p^{\rho}\, \Lambda_{\rho\mu} +\frac{\p_{\mu}\, \Lambda}{D}\,\,,\label{diffs3b} \ee 
 
 \no the expression (\ref{dellstar}) suggests the non local field redefinition:
 
 \be \hmn = \vf_{\mu\nu\rho} + 2\, \frac{\p_{(\mu}\p_{\nu}\tchi_{\rho)}}{\Box} - \frac{3(D-2)}{(D-1)}\,  \frac{\p_{\mu}\p_{\nu}\p_{\rho}(\p\cdot \, \tchi)}{\Box^2}  \, . \label{fredefs3b} \ee
 
 \no After substituting (\ref{fredefs3b}) in $S(a_{D+2},b,c_S)$ once again the non local terms are cancelled out and we end up with a local action whose Lagrangian density is:
 
 \bea {\cal L}(\vf,\tchi) &=& \frac 12 \vf_{\mu\nu\rho}\, \Box \, \vf^{\mu\nu\rho} + \frac 32 (\p^{\mu}\vf_{\mu\nu\rho})^2 + 3\, a_{D+2} \, \p^{\mu}\p^{\nu} \vf_{\mu\nu\rho} \, \vf^{\rho}- \frac 32\, b \, \vf_{\mu}\, \Box \, \vf^{\mu} + \frac 34\, c_S \, (\p^{\mu} \, \vf_{\mu})^2 \nn\\\
 &+& 6\,(a_{D+2}-1)(\p\cdot\p\cdot\vf)\cdot \tchi + \, 6(a_{D+2}-b)\tchi^{\mu}\, \Box \, \vf_{\mu} -6 \left\lbrack b+\frac{D-4}{(D+2)^2}\right\rbrack (\p\cdot \vf)(\p\cdot \tchi) \nn\\
 &+& 6(2 a_{D+2}-1-b)\tchi_{\mu}\, \Box \, \tchi^{\mu} -3[2\, b -1 + (D-1)a_{D+2}](\p\cdot \tchi)^2 \, . \label{l3b} \eea 
 
 \no  After the invertible field redefinition:

\be (\vf_{\mu\nu\rho},\tchi_{\mu})= \left(H_{\mu\nu\rho} + \eta_{(\mu\nu}B_{\rho)} ,  - \frac{H_{\mu}}{2}\right) \, , \label{fredefs3diag} \ee

\no we end up with  the pure  spin-3 Fronsdal model decoupled from the Maxwell  theory, 

\be {\cal L}[a_{D+2},b,c_S] = {\cal L}_F(1,1,1,H) + \frac{3(D+2)^2\, f_{D+2}(a_{D+2},b)}D \left( -\frac 14 F_{\mu\nu}^2(B)\right)  \quad , \label{f+mb} \ee

\no which confirms the particle content displayed in (\ref{a2ks3}) at $a=a_{D+2}$. So once more a non local field redefinition, see (\ref{fredefs3b}) has not changed the particle content. 

As in previous cases, for sake of comparison with the doublet action (\ref{st}), it is convenient to make an invertible change of variables. 
Notice that the gauge transformation of $\tchi$ in (\ref{diffs3b}) is slightly different from (\ref{diffs}). However, 

\be \tchi = \frac 1D \left\lbrack \vf_{\mu} - (D+2) \, \chi_{\mu}\right\rbrack \qquad \to \qquad  \delta\, \chi_{\mu} = \, \p^{\nu}\, \Lambda_{\mu\nu} \, , \label{tchi} \ee

\no After implementing (\ref{tchi}) in (\ref{l3b}) we obtain

\bea {\cal L}[a_{D+2},b,c_S] &=& \frac 12 \vf_{\mu\nu\rho}\, \Box \, \vf^{\mu\nu\rho} + \frac 32 (\p^{\mu}\vf_{\mu\nu\rho})^2
+ 6\, \chi\, (\p\cdot\p\cdot \vf) - 6 \, \chi^{\mu}\, \Box \, \chi_{\mu} +3 \,(\p\cdot \chi)^2 \nn\\ &-& \frac{3[4-b(D+2)^2]}{4\, D^2}\, F_{\mu\nu}^2(\vf-2\, \chi) \,\,. \label{stgs3b} \eea

If we look at the first line of both (\ref{stgs2}) and (\ref{stgs3}) we recognize the spin-2 and spin-3 versions of (\ref{st}) respectively. Since the rest of those actions vanishes at $a=0=b$, we see that is possible to connect the local one field Maxwell-like models of \cite{cf} with the local doublet action stemming from the tensionless limit of string field theory in one  step by means of non local field redefinitions which introduce  a rank-$(s-2)$ symmetric field. It is mentioned in \cite{cf} that this could be done in two steps via  a rank-$(s-1)$ symmetric field whose divergence would be identified with the rank-$(s-2)$ field $\chi$.

The question arises: what about the general case $(a,b)\ne (0,0)$? It turns out that the extra terms beyond (\ref{st}) in (\ref{stgs2}) and (\ref{stgs3})  are gauge invariant on their own. In the spin-2 case this follows from  $\delta\, (\vf-\, 2\chi)=0$ and the invariance of the linearized scalar curvature: $\delta (\p^{\mu}\p^{\nu}\vf_{\mu\nu} - \Box\,\vf)=0$ under (\ref{diffs2}). For $s=3$ the diff symmetry coincides with  the usual $U(1)$ symmetry of Maxwell theory since under (\ref{diffs3}) we have $\delta\, (\vf_{\mu} - 2\, \chi_{\mu})=\p_{\mu}\Lambda $.  Therefore, in both $s=2$ and $s=3$ cases the $(a,b)$ dependent difference with respect to (\ref{st}) is trivial from point of view of the gauge symmetry. However, we should not identify such difference  immediately  with a cohomologically trivial term regarding the BRST charge since, as we have seen,  according to the value  of $(a,b)$ the spin-0 particle may be a ghost instead of the  $a=0=b$ case where it is a physical particle, this applies also to the $s=3$ case regarding the spin-1 particle. Even if we are constrained to a subspace of the two dimensional parameters space $(a,b)$ where the spin-$(s-2)$ particle remains physical it is not possible in general to connect all models with different values of $(a,b)$ via local field redefinitions to $(a,b)=(0,0)$. The turning point for the understanding of the mismatch between (\ref{st}) and (\ref{stgs2}) for $s=2$ and  between (\ref{st}) and (\ref{stgs3}) for $s=3$ for $(a,b)\ne (0,0)$ came up from the special $s=3$ case $a=a_{D+2}$.

We see in (\ref{stgs3b})  that the extra term  (Maxwell term) is once again gauge invariant by itself. However, if $b=a_{D+2}^2$, which implies $c_S=( a_{D+2})^2=b $, we can also go in one (non local) step from the one field T$^*$Diff model $S^*\equiv S[a_{D+2},a_{D+2}^2,a_{D+2}^2]$ directly to the doublet model coming from string theory (\ref{st}). Therefore, $S^*$ must be equivalent to the spin-3 Maxwell-like model $(a,b,c)=(0,0,0)$. Indeed We have been able to prove that there is a nontrivial equivalence between  those models which we have later generalized, proving that any $(a,b,c_S)$ spin-3 model within the physical subspace $f_{D+2} >0 $ is equivalent to the Maxwell-like model $(a,b,c)=(0,0,0)$. A similar equivalence holds also for the $s=2$ case.
It can be shown from the two point amplitude via a simultaneous redefinition of the sources and of the constants of the model or alternatively from the local action  via a local followed by a nonlocal field redefinition as we explain in the next section. 

\section{Equivalence between TDiff models}

In this subsection we prove that the spin-2 $S(a,b)$ and spin-3 $S(a,b,c_S)$ 
 TDiff models, see  (\ref{tds2}) and (\ref{sts}) respectively, are equivalent to the corresponding  Maxwell-like models $S(0,0)$ and $S(0,0,0)$. 

\subsection{Sources redefinition  ($s=2$)}

First, we notice that
 after performing the invertible local field redefinition:   $h_{\mu\nu}\rightarrow  h_{\mu\nu}+ r\, h \,\eta_{\mu\nu}$  in (\ref{tds2}), with $r\neq-1/D$ , called  $r$-shifts henceforth, the form of the TDiff action is preserved with the map $(a,b) \to (a(r),b(r))$ where,
\bea
a(r)&=& a+r(D\;a-2),\label{al}\\
b(r)&=& b (1+r\, D)^2 -2\, (a+1)\, r \,(1+r\, D)+ r^2 (D+2) \quad , \label{bl}\\
f_D [a(r),b(r)]&=& (1+r\, D)^2  \, f_D (a,b) \quad . \label{fdabl} 
\eea

 Since the $r$-shifts are trivial field redefinitions they should not lead to any physical consequence. Due to the source term in (\ref{tds2}), in practice we must compare the model $(a,b,T_{\mu\nu},J)$ to $(a(r),b(r),T_{\mu\nu}(r),J(r))$ where $T_{\mu\nu}(r)=T_{\mu\nu}+r\, \eta_{\mu\nu}T$ and  $J(r)= J + r\, T$.  The conserved source $\tilde{T}_{\mu\nu}$ of the pure spin-2 sector is $r$-shift invariant
  
    \be \tilde{T}_{\mu\nu}(r) = T_{\mu\nu}(r)-\eta_{\mu\nu} J(r) = \tilde{T}_{\mu\nu}\,\,, \ee  
    
\no while $\tilde{J}(r) = (1+r\, D)\tilde{J} $. Thus, from the two point amplitude (\ref{ampldiag}) and the definitions (\ref{tildetj}), recalling (\ref{fdabl}), we show that:
  
   \be \mathcal{A}_2[a,\,b,\,T_{\mu\nu},\,J] = \mathcal{A}_2[a(r),\,b(r),\,T_{\mu\nu}(r),\,J(r)] \quad , \label{Ar} \ee

  \no which confirms the triviality of the $r$-shifts as far as the sources, or equivalently the couplings to the sources, are redefined accordingly.
  
  Starting from an arbitrary pair $(a,b)$ we cannot bring it to $(0,0)$ by performing only the one-parameter $r$-shifts. However, we can do it by two steps by also redefining the sources through an independent real parameter\footnote{The real parameter $s$ should not be confused with the spin.} ($s$-shift):
  
  \be (T_{\mu\nu}(s),J(s)) \equiv (T_{\mu\nu} + s\, \eta_{\mu\nu} J,(1+s)J)    \quad . \label{tc} \ee
  
  Now let us require equivalence of the amplitude of the Maxwell-like model with $s$-shifted sources on one hand with an $(A,B)$ TDiff model  on the other hand, using (\ref{ampldiag}) we have
  
  \be 
 \tilde{T}^{*}_{\mu\nu}(s)\tilde{T}^{\mu\nu}(s)-\frac{|\tilde{T}(s)|^2}{D-2}+\frac{|\tilde{J}(s)|^2}{(D-2)f_{D}(0,0)} =   \tilde{T}^{*}_{\mu\nu}\tilde{T}^{\mu\nu}-\frac{|\tilde{T}|^2}{D-2}+\frac{|\tilde{J}|^2}{(D-2)f_{D}(A,B)}
 \label{equiv1} \ee
 
 \no Similarly to the $r$-shifts, the $s$-shifts do not change the conserved source of the pure spin-2 sector $\tilde{T}_{\mu\nu}(s) = T_{\mu\nu}(s)-\eta_{\mu\nu} J(s) = \tilde{T}_{\mu\nu} $, so we are left only with the $\vert\tilde{J}\vert^2$ terms on both sides of (\ref{equiv1}). Therefore, recalling that the definition of $\tilde{J}$ itself, see (\ref{tildetj}),  depends on the corresponding constant $a$ and using $f_D(0,0)=1$,  the equation (\ref{equiv1}) becomes
 
 \be \vert T + [s(D-2)-2]\, J \vert^2 = \frac{(1-A)^2}{f_{D}(A,B)} \,\Big\vert T + \frac{(A\, D -2)}{1-A} J \Big\vert^2  \quad , \label{equiv2} \ee
 
\no where we have assumed $A\ne 1$ (we comment on that point later). Consequently, see (\ref{fdab}), the equivalence (\ref{equiv2}) requires
  
 \bea  s(D-2) -2 &=& \frac{A\, D -2}{1-A}  \quad\quad\quad \to \quad s= \frac A{1-A} \label{ba2a} \\
  (1-A)^2 &=& f_{D}(A,B) \quad\,\,\,\quad \to \quad B=A^2 \quad , \label{ba2}  \eea

Our final step amounts to show that any pair $(a,b)$ can be brought to $(A,A^2)$ with $A\ne 1$ via the local $r$-shifts. Indeed, defining $(A,B)=(a(r),b(r))$, from (\ref{al}) and (\ref{bl}) it can be shown that $B=A^2$ is equivalent to 

\be \left\lbrack D^2f_D - (a\, D -2)^2\right\rbrack \, r^2 + 2\left\lbrack D\, f_D + (1-a)(a\, D -2)\right\rbrack \, r + f_D-(a-1)^2 = 0 \, , \label{A2B} \ee
 \no where in (\ref{A2B}) it is understood that $f_D=f_D(a,b)$. Moreover, from (\ref{al}), the condition $A\ne 1$ is equivalent to 
 
 \be r \ne \frac{(1-a)}{(a\, D -2)} \quad , \quad {\rm if} \quad a\ne \frac{2}{D} \, , \label{rs2} \ee
 
 \no while it is automatically satisfied at $a=2/D$, since throughout this work $D\ge 3$. 
 
 The two roots of (\ref{A2B}) are given by 
 
 \be r^{\pm}_D = \frac{\pm \sqrt{f_D} + a-1}{(2-a\, D)\mp D\, \sqrt{f_D}} \quad . \label{rpm} \ee 
 
 \no Since a TDiff model requires $f_D \ne 0$, the condition (\ref{rs2}) is always satisfied. Furthermore, if the denominator of one of the roots 
 $r^{\pm}_D$ vanishes, we can always use the other one in order to bring $B\to A^2$, thus avoiding any singularity. Last, we should not worry either about the non invertible $r$-shift $r^{\pm}_D=-1/D$, since it can only happen at $D=2$. 
 
 Thus, we can always bring an arbitrary $(a,b)$ physical ($f_D >0$) TDiff model  to another one $(a(r),a^2(r))$ with $a(r)\ne 1$ via an  $r$-shift given in (\ref{rpm}). Since the $r$-shift does not change the amplitude as we have seen in (\ref{Ar}), we end up with the following equivalence of amplitudes:
  
 \be \mathcal{A}_2[0,\,0,\,T_{\mu\nu}(r,s),\,J(r,s)] = \mathcal{A}_2[a(r),\,a^2(r),\,T_{\mu\nu}(r),\,J(r)] = \mathcal{A}_2[a,\,b,\,T_{\mu\nu},\,J] \quad , \label{Ar2} \ee
  
  \no where:
  
  \bea (T_{\mu\nu}(r),J(r)) &=& (T_{\mu\nu} + r\, \eta_{\mu\nu}\, T\, , \, J + r \, T) \quad , \label{tr}\\(T_{\mu\nu}(s,r),J(s,r)) &=& (T_{\mu\nu}(r) + s\, \eta_{\mu\nu} J(r)\,,\,(1+s)J(r))    \quad . \label{tcr} \eea
  
  In summary, the equality (\ref{Ar2}) establishes that the physics of the spin-2 TDiff model $S(a,b,T_{\mu\nu},J)$ given in (\ref{tds2}) with $(a,b)$ such that $f_D >0$ is equivalent to the physics of the Maxwell-like model $S(0,0,T_{\mu\nu}(s,r),J(s,r))$. The new sources are linear combinations of old ones as defined in (\ref{tr}) and (\ref{tcr}) with $r=r_D^{\pm}$ and $s=s(r_D^{\pm})$ as given in (\ref{rpm}) and (\ref{ba2a}) respectively and recalling that $A=a(r_{D}^{\pm})$ according to (\ref{al}). Alternatively, in terms of the original sources $(T_{\mu\nu},J)$ used in the $(a,b)$ model  one can say that the coupling of the Maxwell-like model $(0,0)$ must be modified in order to furnish the same amplitude of the $(a,b)$ model. Since the relationship between $(T_{\mu\nu}(s,r),J(s,r))$ and $(T_{\mu\nu},J)$
  is always invertible we can equivalently consider the Maxwell-like model as simply coupled to sources via $h_{\mu\nu}T^{\mu\nu}$ and show its equivalence to an $(a,b)$ physical TDiff model with a reverse modified coupling to those sources. 
  
  Moreover, since each $(a,b)$ model  with $f_D(a,b)>0$ is equivalent to the Maxwell-like model, it follows that any  $(a,b)$ physical model must be equivalent to any other  $(\tilde{a},\tilde{b})$ with $f_D(\tilde{a},\tilde{b})>0$, as far as we modify the couplings to the sources conveniently. Completely similar conclusions will be drawn in the next subsection about the spin-3 case.

  \subsection{Sources redefinition  ($s=3$)}

Since the spin-3 equivalence is quite close to the spin-2 one, we can already 
start at the spin-3 generalization of (\ref{equiv1}). Namely,  let us require equivalence of the amplitude of the Maxwell-like model $S(0,0,0)$ with $s$-shifted sources $(T_{\mu\nu\rho}(s),J_{\mu}(s))$ on one hand with a TDiff model $S(A,B,c_S(A,B))$ with usual sources $(T_{\mu\nu\rho},J_{\mu})$  on the other hand, where $c_S(A,B)=3\, A^2 - 2\, B$. We require, see (\ref{a2ks3}),

 \be 
\left[\tilde{T}^*_{\mu\nu\rho}(s)\tilde{T}^{\mu\nu\rho}(s)-\frac 3D \, |\tilde{T}_{\mu}(s)|^2 \, + \, \frac{3\,|\tilde{J}^T_{\mu}(s)|^2 }{D\, f_{D+2}(0,0)}\right] = \left[\tilde{T}^*_{\mu\nu\rho}\tilde{T}^{\mu\nu\rho}-\frac 3D \, |\tilde{T}_{\mu}|^2 \, + \, \frac{3\,|\tilde{J}^T_{\mu}|^2 }{D\, f_{D+2}(A,B)}\right] \\ \label{equiv3}.
\ee
 
\no The quantities on the left side of (\ref{equiv3}) are given, see (\ref{jtil}) with $a=0$, by

\bea  \tilde{T}_{\mu\nu\rho}(s) &= &  T_{\mu\nu\rho}(s)\, - \eta_{(\mu\nu}\, J_{\rho)}(s) \quad , \label{ttils3} \\ T_{\mu\nu\rho}(s) &= &  T_{\mu\nu\rho} \, +s\, \eta_{(\mu\nu}\, J_{\rho)} \quad , \label{ts3} \\
\tilde{J}^T_{\mu}(s) &\equiv & 2\,  J_{\mu}(s) \, -\, T_{\mu}(s) \quad , \label{jtils3} \\ f_{D+2}(0,0)&=& 1 \label{fd20} \eea 
  
\no where $J_{\rho}(s)$ is obtained as follows. Using (\ref{can1})
we have 

\bea \p^{\mu}T_{\mu\nu\rho}(s) &=& \p^{\mu}T_{\mu\nu\rho} + s\, \eta_{\nu\rho} \p \cdot J + s \, (\p_{\nu}J_{\rho} + \p_{\rho} J_{\nu}) \nn\\
&=& (1+s)\,(\p_{\nu}J_{\rho} + \p_{\rho} J_{\nu}) + (s+\kappa)\, \eta_{\nu\rho} \p \cdot J \label{dts3}  \eea

\no with 

\be \kappa = \frac{-2\,k}{(1+k\, D)} = \frac{A}{(A-1)} \, , \label{kappa} \ee

\no Since $a=0 $ implies $k=0$, see (\ref{cstar}), there should be no term proportional to $\eta_{\mu\nu}$ on the right side of (\ref{dts3}). Therefore,

\be s = - \kappa = \frac A{1-A} \label{s3} \ee

\no which leads to 

\bea \p^{\mu}T_{\mu\nu\rho}(s) &=& \p_{\nu}J_{\rho}(s) + \p_{\rho} J_{\nu}(s)  \, , \label{ts3} \\ J_{\mu}(s) &\equiv & (1+s)\,  J_{\mu}  = (1-\kappa)\,  J_{\mu} \quad . \label{js3} \eea

\no From the trace of (\ref{ts3}) we check that $\p^{\mu}\tilde{J}^T_{\mu}(s) =0$. Moreover, similarly to the spin-2 case, since the conserved source of the pure spin-3 sector is $s$-shift invariant $ \tilde{T}_{\mu\nu\rho}(s)=\tilde{T}_{\mu\nu\rho}$, the equivalence (\ref{equiv3}), supposing $A\ne 1$,  amounts to 

\be \vert T_{\mu}(s) - 2\, J_{\mu}(s) \vert^2 = \vert T_{\mu} - (2 + \kappa \, D)\, J_{\mu} \vert ^2 = \frac{(A-1)^2}{f_{D+2}(A,B)} \vert T_{\mu} - \frac{[A(D+2)-2]}{A-1} J_{\mu} \vert^2 \,\,\, . \ee

\no It turns out from (\ref{s3}) that $2 + \kappa \, D=[A(D+2)-2]/(A-1)$, thus we only need 

\be f_{D+2}(A,B) = D(A^2-B) + (A-1)^2 = (A-1)^2 \quad \to \quad B=A^2\,\,\, . \label{ba23} \ee

\no It is the same restriction of the spin-2 case, see (\ref{ba2}), so we proceed similarly. Under spin-3 invertible $r$-shifts  in (\ref{sabc}), $h_{\mu\nu\rho}\rightarrow  h_{\mu\nu\rho}+ r\, \eta_{(\mu\nu}\, h_{\rho)}$, with $r\neq-1/(D+2)$, we have $(a,b,c) \to (a(r),b(r),c(r))$ such that,

\bea
a(r)&=& a+r(a\, D + 2\, a-2),\label{a3}\\
b(r)&=& b [1+ r \,(D + 2)]^2 -2\, (a+1)\, r\, [1+ r \,(D + 2)]+ r^2 (D+4) \quad , \label{b3}\\
c(r) &=& c [1+ r \,(D + 2)]^2 -8\, a\, r^2(D+2)+4\, r(1-2\,  a) + 2\, r^2(D+6)
\, , \label{c3} \eea

\no which leads to \footnote{One can also show that the condition $c=c_S$ is invariant under $r$-shifts since: $[c(r)-3\, a(r)^2 + 2\, b(r)] = [1+r(D+2)]^2 (c-3\, a^2 + 2\, b)$.}  $ f_{D+2} [a(r),b(r)] = [1+r\, (D+2)]^2\, f_{D+2} (a,b)$. 

 If, as in the spin-2 case, we identify $(A,B)=(a(r),b(r))$, since (\ref{a3}) and (\ref{b3}) correspond to the expressions (\ref{al}) and (\ref{bl}) of the spin-2 case with the replacement $D\to D+2$,  the condition $B=A^2$ leads to the $r$-shifts: 

\be r^{\pm}_{D+2} = \frac{\pm \sqrt{f_{D+2}} + a-1}{[2-a\, (D+2)]\mp (D+2)\, \sqrt{f_{D+2}}} \quad . \label{rpm3} \ee 

\no Notice from (\ref{a3}) and (\ref{rpm3}) that  $A=a(r)\ne 1$ is assured by $f_{D+2}\ne 0$ and we never have non invertible $r$-shifts, i.e., $r^{\pm}_{D+2}\ne-1/(D+2)$, so the equivalence is established also for $s=3$. 

In conclusion, we have shown that the amplitude obtained
from $S(a,b,c_S,T_{\mu\nu\rho},J_{\mu})$ given in (\ref{sts}) is equivalent to the amplitude obtained from $S(0,0,0,T_{\mu\nu\rho}(r,s),J_{\mu}(r,s)) $, which is the spin-3 Maxwell-like TDiff model with redefined sources, where 

 \bea (T_{\mu\nu\rho}(r),J_{\mu}(r)) &=& (T_{\mu\nu\rho} + r\, \eta_{(\mu\nu}\, T_{\rho)}\, , \, J_{\mu} + r \, T_{\mu}) \quad , \label{tr3}\\(T_{\mu\nu\rho}(s,r),J_{\mu}(s,r)) &=& (T_{\mu\nu\rho}(r) + s\, \eta_{(\mu\nu} J_{\rho)}, \, (1+s)J_{\mu}(r))    \quad , \label{tcr3} \eea

\no with $r=r_{D+2}^{\pm}$ and $s=s(r_{D+2}^{\pm})$ given by (\ref{rpm3}) and (\ref{s3}) respectively, while $A=a(r_{D+2}^{\pm})$ is given by (\ref{a3}). 

As in the spin-2 case, since $(T_{\mu\nu\rho}(r,s),J_{\mu}(r,s))$ are linear combinations of the original sources $(T_{\mu\nu\rho},J_{\mu})$, we may say that the model $S(a,b,c)$ with $f_{D+2}(a,b)>0$ usually coupled to the sources via $h_{\mu\nu\rho}T^{\mu\nu\rho}$  is equivalent to the Maxwell-like model $S(0,0,0)$ with redefined couplings via $h_{\mu\nu\rho}T^{\mu\nu\rho}(s,r)$. 

In the next subsection we present an alternative derivation of equivalence based on non local field redefinitions for both spin-2 and spin-3 cases. 

\subsection{Equivalence via nonlocal field redefinitions} 

Let us start with the simpler spin-2 case. Let us suppose that we  begin with an arbitrary $S(a,b)$ TDiff model in (\ref{tds2}), so far without sources, which is moved into $S(A,B)=S(a(r),b(r))$ after an $r$-shift. Next, performing the non local field redefinition:

\be h_{\mu\nu} \to h_{\mu\nu} + s\, \frac{\p_{\mu}\p_{\nu}h}{\Box} \quad  , \label{nls2} \ee

\no the local form of the TDiff action is preserved such that  $ S(A,B) \to S(a(s,r),b(s,r)) $. In order to arrive at the Maxwell like theory we require:

\bea a(s,r) &\equiv &  1+(A-1)(1+s) = 0 \quad , \label{as} \\
b(s,r) &\equiv& 1- 2 \, (A-1)(1+s)\, s + (B-1)(1+s)^2 = 0  \quad . \label{bs} \eea

\no The solution of (\ref{as}) and (\ref{bs}) only exists if $A\ne 1$ and leads, in full agreement with subsection 4.1, to 

\be s= \frac A{1-A} \quad ; \quad B=A^2   \quad , \label{sol2} \ee 

\no Since we already know that $b(r)=a^2(r)$  can be implemented via (\ref{rpm}), we conclude that such $r$-shift followed by the nonlocal redefinition (\ref{nls2}) leads us to the Maxwell-like model, 

\be S(a,b)  \Rightarrow  S(a(r),b(r)) \Rightarrow  S(0,0) \quad . \label{equiv4} \ee

Several comments are necessary. First, notice that the solution (\ref{sol2}) is such that we never have $s=-1$ which would make (\ref{nls2}) non  invertible even if we neglected the kernel of $\Box$. More severe is the fact that we are dealing with a massless theory, so we can not rigorously justify the nonlocal redefinition anyway. It is however, remarkable that if we take into account the source term, the redefinition (\ref{nls2}) is in full agreement with the local source redefinition of the last section. Namely, disregarding total derivatives and recalling 
$\p^{\mu}T_{\mu\nu} = \p_{\nu}J$, 

\be h^{\mu\nu}T_{\mu\nu} \to \left( h^{\mu\nu} + s \, \frac{\p^{\mu}\p^{\nu}h}{\Box}\right) T_{\mu\nu} = h^{\mu\nu} \left( T_{\mu\nu} + s\, \eta_{\mu\nu}\, J\right) =   h^{\mu\nu} \, T_{\mu\nu}(s) \quad . \label{sources2} \ee

\no As in subsection 5.1, we do not need to worry about the restriction $A\ne 1$ since $f_D \ne 0$ assures $r^{\pm}_D \ne (1-a)/(a\, D -2) $ if $a\ne 2/D$ while at $a= 2/D$ the condition $a(r)\ne 1$ holds automatically, see (\ref{al}). 

Regarding the spin-3 case, we use the generalization of (\ref{sources2}) as a short cut, but now we have more freedom to choose the nonlocal field redefinition, namely, 

\be h_{\mu\nu\rho} \to h_{\mu\nu\rho} + s_1\, \frac{\eta_{(\mu\nu}\p_{\rho )}\, \p \cdot h}{\Box} + s_2\, \frac{\p_{(\mu}\p_{\nu}h_{\rho )}}{\Box } + s_3\, \frac{\p_{\mu}\p_{\nu}\p_{\rho} \, \p\cdot h}{\Box^2} \, , \label{nls3} \ee

\no Using (\ref{nls3}) and (\ref{can1}), after integrations by parts, we require, see (\ref{ts3}),

\bea h^{\mu\nu\rho}\, T_{\mu\nu\rho} &\to &  h^{\mu\nu\rho}\,[T_{\mu\nu\rho} + s_2\, \eta_{(\mu\nu}\, J_{\rho)} ] - \left[ 3(2+\kappa\, D)\, s_1 + 3\, s_2 (1+\kappa) + s_3\, (2+\kappa) \right] \frac{(\p\cdot h)(\p\cdot J)}{\Box} \nn\\ &=&   h^{\mu\nu\rho}\,[T_{\mu\nu\rho} - \kappa\, \eta_{(\mu\nu}\, J_{\rho)} ] \,\,\,.\label{require} \eea

\no Assuming $A\ne 1$ without loss of generality and further supposing $A\ne 2/(D+2)$, we come to that case later, the requirement (\ref{require}) leads to 

\bea s_2 &=& -\kappa = \frac A{1-A} \quad , \label{s2} \\
s_1 &=& \frac{A(2\, A -1)}{(A-1)[A(D+2)-2]}- \frac{s_3 \, (A-2/3)}{(A-1)[A(D+2)-2]}  \quad . \label{s1} \eea

\no We have explicitly checked by means of the software Mathematica with help of the package {\it xAct} \cite{teake} that (\ref{nls3}) with
(\ref{s2}) and (\ref{s1}) in (\ref{sabc}) with $(a,b,c)=(A,B,c_S)$  indeed leads to the Maxwell-like model $(0,0,0)$ if $B=A^2$ and $c_S =3\, A^2- 2\, B \Leftrightarrow c_S=A^2 $ for arbitrary values of $s_3$. If $A=2/(D+2)$ the condition (\ref{require}) leads to the same value of $s_2$ in (\ref{s2}) but now $s_1$ becomes a free parameter while $s_3= -3(D-2)/[D(D-1)]$. There is one exception in the parameters space which must be mentioned. From (\ref{nls3}) we have
 $\p \cdot h \to   U \, \p \cdot h$ where $U= s_1(D+2)-3\kappa + s_3 +2 $. An invertible field redefinition requires $U\ne 0$ which leads to  $s_3 \ne 3(A\, D -1)/(A-1) $ if $A\ne 2/(D+2)$ and $s_1\ne -1/(D-1)$ if $A= 2/(D+2)$ . The new trace $h_{\mu}$ can also be obtained from the old one if $\kappa \ne 1$ which is always the case. 
 

 In summary, in both spin-2 and spin-3 cases the Maxwell-like model can be quickly obtained from an arbitrary TDiff model via an $r$-shift followed by a nonlocal field redefinition, though we have not been able to rigorously justify the latter ones. In the spin-3 case the nonlocal redefinition is not unique.

At this point a comment is in order. According to \cite{francia_irred} and \cite{francia_triplets} where non local higher spin irreducible and reducible free models respectively, are obtained after integration over Stueckelberg-like fields, it is natural
to think of turning our non local transformations into local ones with the help of the Stueckelberg-like field $\chi$ of rank-$(s-2)$. Indeed, the authors of \cite{cubic} have been able to give an alternative derivation of a self-interacting vertex for spin-2 fields in the Maxwell-like TDiff model $(a,b)=(0,0)$ by turning a previously non local non linear field redefinition involving the field $h_{\mu\nu}$ into a local one with the help of the Stueckelberg-like scalar field.  In the spin-2 case we have tried the following local redefinition 
 
 \bea h_{\mu\nu} &=& H_{\mu\nu} + \eta_{\mu\nu}(p\, \tilde{\chi} + q\, H) \label{fr1} \\
 \chi &=& \tilde{\chi} + t\, H \label{fr1} \eea
 
 \no in order to get rid of the second line of (\ref{stgs2}) and turn it into the doublet string action (\ref{st}). It turns out that it only works if $b=b_S(a,D)\equiv a[4-a(D-2)]/4$, in this case $(p,t)= (-2\, q,0)$ and $q=a/(2-a\, D)$. However, the spin-2 TDiff model $(a,b_S(a,D))$ is nothing but an $r$-shifted version, with $r=-a/2$, of the Maxwell-like model $a=b=0$. If we perform a similar field redefinition in the spin-3 case in order to get rid of the Maxwell term in (\ref{stgs3})
 it is required that $b=b_S(a,D+2)\equiv a(4-a\, D)/4$ which is such that $f=0$, see (\ref{g}). This corresponds again to an $r$-shift of the spin-3 Maxwell-like TDiff model. Since $r$-shifts are local field redefinitions we make no progress, regarding the nonlocal $s$-shifts, by using the Stueckelberg-like field.

 \section{Conclusion}
 
 We have introduced a general spin-3 model $S(a,b,c)$ in $D$ dimensions ($D\ge 3$) in (\ref{sabc}), defined in terms of a rank-3 symmetric tensor and three real parameters. Requiring invariance under a generalized version of transverse diffeomorphisms (TDiff) suggested in (\ref{gtdiff}) and (\ref{cstar}) for $a\ne 2/(D+2)$ and in (\ref{lstar}) and (\ref{vector2}) for $a= 2/(D+2)$, we have found the condition $c=c_S\equiv 3\, a^2 - 2\, b$ which seems to be rather universal and holds also for the pure spin-3 (irreducible) Fronsdal \cite{fronsdal} and Skvortsov-Vasiliev \cite{sv} models. 

The generalized TDiff model $S(a,b,c_S)$ is reducible and  describes a massless spin-3 plus a massless spin-1 particle. The latter may be physical or not according to the values of $(a,b)$. In order to investigate the particle content of the spin-3 TDiff model we have used a basis of non local differential operators acting on symmetric rank-3 tensors, see \cite{SchimidtElias} and the appendix, allowing us to obtain the gauge invariant two point amplitude (\ref{a2ks3}). 

In section 4 we have compared the spin-2 and spin-3 TDiff models  with the ``doublet'' action (\ref{st}) coming from the tensionless limit of the open bosonic string field theory in the free case, see \cite{st}, after algebraically integrating over one of the fields of the triplet action, see \cite{fscqg} and the earlier literature \cite{ouvry,beng,ht}. The doublet model (\ref{st}) describes massless particles of spin $s,s-2,s-4,\cdots, 0$ or $1$ according to even or odd $s$ respectively and it is invariant under unconstrained diffeomorphisms (Diff).

In order to compare unconstrained theories like (\ref{st}) with constrained ones it is natural to use Stueckelberg fields, see e.g. \cite{sagnotti_notes,cf}.  We
have developed a specific method of introducing Stueckelberg fields and compared (\ref{st}) with our singlet (one field) TDiff models. The method introduces a rank-$(s-2)$ Stueckelberg-like field ($\chi$) via non local field redefinitions, see (\ref{fredefs2}),(\ref{fredfs3}) and (\ref{fredefs3b}). Although non local, we have shown, after a simple diagonalisation\footnote{For Maxwell-like models $(a,b)=(0,0)$ there is a non local diagonalisation in \cite{cf} for arbitrary integer spins. However, in \cite{cf} the original (tracefull) rank-$s$ field is decomposed in traceless (irreducible) lower rank components and the TDiff symmetry is not enlarged to Diff.}, that the redefinitions do not change the spectrum of the original singlets. Moreover they extend the TDiff symmetry to unconstrained Diff which are more natural in String theory. For sake of comparison we have written our models in a convenient way in (\ref{stgs2}) for spin-2 and in (\ref{stgs3}) and (\ref{stgs3b}) for spin-3 with $a\ne 2/(D+2)$ and $a=2/(D+2)$ respectively. For the actions displayed in (\ref{stgs2}) and (\ref{stgs3}) we only  have apparently  full agreement with the doublet model (\ref{st}) at vanishing values of our free parameters ($a=0=b$). That point corresponds to the Maxwell-like models introduced in \cite{cf} for arbitrary integer spins.

Regarding the point $a=a_{D+2}=2/(D+2)$ with  $c=c_S=3\, (a_{D+2})^2 - 2 \, b$, the parameter $b$ is the only free one, see (\ref{stgs3b}), the model explicitly coincides with the doublet theory only at $b=(a_{D+2})^2$ which is not of the Maxwell-like type. So we conclude that the spin-3 Maxwell-like model $S(0,0,0)$ must be equivalent  to the model $S(a_{D+2},a_{D+2}^2,c_s)$. However, there is no local field redefinition which might bring one model to the other one, since $a=2/(D+2)$ is a fixed point of the local redefinitions that we have called $r$-shifts, see (\ref{a3}). This point and the apparent discrepancies between the TDiff models and the doublet action has led us to prove  in section 5 a non trivial equivalence among ghost free TDiff models of spin-2 and spin-3.

We have shown that any spin-2 TDiff model $S(a,b,T_{\mu\nu},J)$ such that $f_D(a,b)>0$, see (\ref{tds2}) and (\ref{fdab}), leads to the same amplitudes of the Maxwell-like model $S(0,0,T_{\mu\nu}(s,r),J(s,r))$ where the modified sources are linear combinations of the original ones, see (\ref{tr}) and (\ref{tcr}) with the real constants $(r,s)=(r_{D}^{\pm},s(r_{D}^{\pm}))$ given in (\ref{rpm}) and (\ref{ba2a}), see subsections 5.1. For the spin-3 case we have similarly proven the physical equivalence of TDiff models $S(a,b,c_S,T_{\mu\nu\rho},J_{\mu})$ such that $f_{D+2}(a,b) >0$ with the spin-3 Maxwell-like model $S(0,0,0,T_{\mu\nu\rho}(r,s),J_{\mu}(r,s))$ with modified sources given in (\ref{tr3}) and (\ref{tcr3}), see subsection 5.2.  Alternatively, we have shown that the equivalence follows from a local ($r$-shift) and a nonlocal ($s$-shift) field redefinition, see subsection 5.3 for both spin-2 and spin-3. So we end up with no physical discrepancies between ghost free spin-2 and spin-3 TDiff models and the doublet action coming from string theory.

In order to get a better understanding of the issues investigated here we clearly need to go to even higher spins ($s>3$ ) especially because of the double traceless condition which appears in Fronsdal models for $s\ge 4$. Indeed, we are currently investigating the spin-4 case and the generalization of our results for (A)dS backgrounds as well as for mixed symmetry tensors as in \cite{cf}. Notice that the non commutativity of the covariant derivatives in the curved space generalization of the nonlocal field redefinition (\ref{nls3}) may be an obstacle to a curved space equivalence. Moreover, the search for self-interacting vertices  for generalized TDiff models, see \cite{cubic} for the Maxwell-like case, is a fundamental issue to be investigated in order to confirm whether the equivalence found here persists at interacting level.

Finally, it is worth mentioning that our method of undoing constraints on gauge parameters via the introduction of Stueckelberg-like fields through field redefinitions may find interesting applications in other gauge field theories.

\section{Acknowledgements}

The work of D.D. is partially supported by CNPq  (grant 306380/2017-0). RSB and BSM are supported by CAPES. BSM had been supported by FAPESP (2023/10876-2 ) during this work.

\section{Appendix: Projection operators} 

We can define projection operators acting on vector fields according to their longitudinal and transverse components, 
\be
\omega_{\mu\nu}=\frac{\del_{\mu}\del_{\nu}}{\square} \quad ; \quad
\theta_{\mu\nu}=\eta_{\mu\nu}-\omega_{\mu\nu}.
\label{omegatheta}
\ee
They satisfy the algebra and the closure relation below, 
\be \theta_{\mu\alpha}\theta^{\alpha\nu}=\theta_{\mu}^{\nu} \qquad \quad ; \quad \omega_{\mu\alpha}\omega^{\alpha\nu}=\omega_{\mu}^{\nu} \quad ; \quad \omega_{\mu\nu}\theta^{\nu\alpha}=0 \quad ; \quad \delta^{\mu}_{\nu}=\theta^{\mu}_{\nu}+\omega^{\mu}_{\nu}, \label{completeza}\ee

From the vector projection operators we can construct a basis for operators acting on rank-3 tensors, following \cite{SchimidtElias}, we have: \bea
(P^{(3)}_{11})^{\mu\nu\rho}_{\alpha\beta\gamma} &=&  \theta^{(\mu}_{(\alpha} \theta^\nu_\beta \theta^{\rho)}_{\gamma)} - (P^{(1)}_{11})^{\mu\nu\rho}_{\alpha\beta\gamma},\label{first} \\
(P^{(2)}_{11})^{\mu\nu\rho}_{\alpha\beta\gamma} &=&  3\, \theta^{(\mu}_{(\alpha} \theta^\nu_\beta \omega^{\rho)}_{\gamma)} - (P^{(0)}_{11})^{\mu\nu\rho}_{\alpha\beta\gamma},  \\
(P^{(1)}_{11})^{\mu\nu\rho}_{\alpha\beta\gamma} &=&  \frac{3}{(D+1)}  \theta^{(\mu\nu} \theta_{(\alpha\beta}  \theta^{\rho)}_{\gamma)},\\
(P^{(1)}_{22})^{\mu\nu\rho}_{\alpha\beta\gamma} &=&  3\, \theta^{(\mu}_{(\alpha} \omega^\nu_\beta \omega^{\rho)}_{\gamma)},  \\
(P^{(0)}_{11})^{\mu\nu\rho}_{\alpha\beta\gamma} &=&  \frac{3}{(D-1)}  \theta^{(\mu\nu} \theta_{(\alpha\beta}  \omega^{\rho)}_{\gamma)}, \label{p110} \\
(P^{(0)}_{22})^{\mu\nu\rho}_{\alpha\beta\gamma} &=&   \omega_{\alpha\beta}
\omega^{\mu\nu} \omega^{\rho}_{\gamma} \label{p220} \eea

\no Notice that, here the parenthesis means normalized symmetrization, taking for example the first term in (\ref{first}) we have: \be \theta^{(\mu}_{(\alpha} \theta^\nu_\beta \theta^{\rho)}_{\gamma)}=\frac{1}{6}(
\theta^{\mu}_{\alpha} \theta^\nu_\beta \theta^{\rho}_{\gamma}+
\theta^{\rho}_{\alpha} \theta^\nu_\beta \theta^{\mu}_{\gamma}+
\theta^{\nu}_{\alpha} \theta^\mu_\beta \theta^{\rho}_{\gamma}+
\theta^{\rho}_{\alpha} \theta^\mu_\beta \theta^{\nu}_{\gamma}+
\theta^{\nu}_{\alpha} \theta^\rho_\beta \theta^{\mu}_{\gamma}+
\theta^{\mu}_{\alpha} \theta^\rho_\beta \theta^{\nu}_{\gamma}).\ee   
The totally symmetric identity operator is represented by $\mathbbm{1}$
and is given by: \bea \mathbbm{1}^{\mu\nu\rho}_{\alpha\beta\gamma} = 
\delta^{(\mu}_{(\alpha} \delta^\nu_\beta \delta^{\rho)}_{\gamma)}. \label{id3} \eea 

\no Finally, the transition operators $P^{(s)}_{{ij}}$
are given by: 
\bea
(P^{(1)}_{{12}})^{\mu\nu\rho}_{\alpha\beta\gamma} &=&  \frac{3}{\sqrt{(D+1)}}   \theta_{(\alpha\beta} \theta^{(\mu}_{\gamma)} \omega^{\nu\rho)},  \\
(P^{(1)}_{{21}})^{\mu\nu\rho}_{\alpha\beta\gamma} &=&  \frac{3}{\sqrt{(D+1)}}   \theta^{(\mu\nu} \theta^{\rho)}_{(\alpha} \omega_{\beta\gamma)},  \\
(P^{(0)}_{{12}})^{\mu\nu\rho}_{\alpha\beta\gamma} &=&  \frac{3}{\sqrt{3(D-1)}}   \theta_{(\alpha\beta} \omega^{(\mu\nu} \omega^{\rho)}_{\gamma)},   \\
(P^{(0)}_{{21}})^{\mu\nu\rho}_{\alpha\beta\gamma} &=& 
\frac{3}{\sqrt{3(D-1)}}   \theta^{(\mu\nu} \omega_{(\alpha\beta} \omega^{\rho)}_{\gamma)}. \eea \no


\begin{thebibliography}{}
	
	\bibitem{snow}
	X.~Bekaert, N.~Boulanger, A.~Campoleoni, M.~Chiodaroli, D.~Francia, M.~Grigoriev, E.~Sezgin, and E.~Skvortsov,
	``Snowmass White Paper: Higher Spin Gravity and Higher Spin Symmetry,''
	[arXiv:2205.01567 [hep-th]].
	
	\bibitem{rt}
	R.~Rahman and M.~Taronna,
	``From Higher Spins to Strings: A Primer,''
	[arXiv:1512.07932 [hep-th]].
	
	\bibitem{sagnotti_notes}
	A.~Sagnotti, 
	``Notes on Strings and Higher Spins,''
	J. Phys. \textbf{A46}, 214006 (2013),
	[arXiv:1112.4285 [hep-th]].
	
	\bibitem{st_r}
	A.~Sagnotti and M.~Taronna,
	``String Lessons for Higher-Spin Interactions,''
	Nucl. Phys. \textbf{B842}, 299–361 (2011),
	[arXiv:1006.5242 [hep-th]].
	
	\bibitem{book}
	A.~Bengtsson,
	``Higher Spin Field Theory, Vol 1+2,''
	De Gruyter, 2023.
	
	\bibitem{porrati}
	M.~Porrati,
	``Universal Limits on Massless High-Spin Particles,''
	Phys. Rev. D \textbf{78}, 065016 (2008),
	doi:10.1103/PhysRevD.78.065016,
	[arXiv:0804.4672 [hep-th]].
	
	\bibitem{bbs}
	X.~Bekaert, N.~Boulanger, and P.~Sundell,
	``How Higher-Spin Gravity Surpasses the Spin-Two Barrier: No-Go Theorems Versus Yes-Go Examples,''
	Rev. Mod. Phys. \textbf{84}, 987–1009 (2012).
	
	\bibitem{fronsdal}
	C.~Fronsdal,
	``Massless Fields with Integer Spin,''
	Phys. Rev. D \textbf{18}, 3624 (1978),
	doi:10.1103/PhysRevD.18.3624.
	
	\bibitem{sv}
	E.~Skvortsov and M.~Vasiliev,
	``Transverse Invariant Higher Spin Fields,''
	Phys. Lett. \textbf{B664}, 301 (2008),
	[arXiv:0701278 [hep-th]].
	
	\bibitem{van}
	J.~J.~van der Bij, H.~van Dam, and Y.~J.~Ng,
	``The Exchange of Massless Spin-Two Particles,''
	Physica \textbf{116A}, 307-320 (1982).
	
	\bibitem{blas}
	E.~Alvarez, D.~Blas, J.~Garriga, and E.~Verdaguer,
	``Transverse Fierz–Pauli Symmetry,''
	Nucl. Phys. \textbf{B756}, 148 (2006);
	D.~Blas,
	``Aspects of Infrared Modifications of Gravity,''
	PhD Thesis, University of Barcelona, 
	[arXiv:0809.3744].
	
	\bibitem{cf}
	A.~Campoleoni and D.~Francia,
	``JHEP,'' 03 (2013) 168,
	[arXiv:1206.5877 [hep-th]].
	
	\bibitem{ouvry}
	S.~Ouvry and J.~Stern,
	``Phys. Lett.,'' \textbf{B177}, 335 (1986).
	
	\bibitem{beng}
	A.~K.~H.~Bengtsson,
	``Phys. Lett.,'' \textbf{B182}, 321 (1986).
	
	\bibitem{ht}
	M.~Henneaux and C.~Teitelboim,
	``Quantum Mechanics of Fundamental Systems,'' Vol. 2, Plenum Press, New York, 1988.

    	\bibitem{st}
	A.~Sagnotti and M.~Tsulaia,
	``On Higher Spins and the Tensionless Limit of String Theory,''
	Nucl. Phys. \textbf{B682}, 83 (2004),
	[arXiv:hep-th/0311257].
    
	\bibitem{rr}
	D.~Dalmazi and R.~R.~L.~d.~Santos,
	``Eur. Phys. J. C,'' \textbf{81} (2021) no.6, 547,
	doi:10.1140/epjc/s10052-021-09297-0,
	[arXiv:2010.12051 [hep-th]].
	
	\bibitem{masterrr}
	Lino dos Santos, R.~R.,
	Master Thesis, São Paulo State University at Guaratinguetá,
	Available at: https://repositorio.unesp.br/handle/11449/193174.
	
	\bibitem{fscqg}
	D.~Francia and A.~Sagnotti,
	``On the Geometry of Higher Spin Gauge Fields,''
	Class. Quant. Grav. \textbf{20}, S473 (2003),
	[arXiv:0212185 [hep-th]].
	
	\bibitem{teake}
	T.~Nutma,
	``xTras: A Field-Theory Inspired xAct Package for Mathematica,''
	Comput. Phys. Commun. \textbf{185}, 1719-1738 (2014),
	doi:10.1016/j.cpc.2014.02.006,
	[arXiv:1308.3493 [cs.SC]].
	
	
	\bibitem{francia_irred} D. Francia, ``On the relation between local and geometric Lagrangians for higher spins'',
	J. Phys. Conf. Ser. 222 (2010) 012002 [arXiv:1001.3854].
	
	\bibitem{francia_triplets} D. Francia, ``String theory triplets and higher-spin curvatures'', Phys. Lett. B 690 (2010) 90
	[arXiv:1001.5003].

	\bibitem{cubic}
	D.~Francia, G.~L.~Monaco, and K.~Mkrtchyan,
	``Cubic Interactions of Maxwell-Like Higher Spins,''
	JHEP \textbf{04}, 068 (2017),
	[arXiv:1611.00292 [hep-th]].

    
    	\bibitem{SchimidtElias}
	E.~L.~Mendonça and R.~Schimidt Bittencourt,
	``Unitarity of Singh-Hagen Model in D Dimensions,''
	Adv. High Energy Phys. \textbf{2020}, 8425745 (2020),
	doi:10.1155/2020/8425745,
	[arXiv:1902.05118 [hep-th]].

	



\end{thebibliography}
\end{document}